\documentclass[11pt,a4paper]{article}

\usepackage{amsfonts}
\usepackage{amssymb}
\usepackage{graphicx}
\usepackage{amsmath}
\usepackage[dvipdfm]{hyperref}
\usepackage{enumerate}

\RequirePackage{mathrsfs} \RequirePackage[sc]{mathpazo}
\RequirePackage{wasysym} \RequirePackage{setspace}
\input{tcilatex}

\begin{document}

\title{On Thermodynamics of AdS Black Holes in Arbitrary Dimensions}
\author{A. Belhaj$^{1}$, M. Chabab $^{2}$, H. El Moumni $^{2}$, M. B. Sedra $%
^{3}$ \\
\\
{\small $^{1}$Centre of Physics and Mathematics, CPM-CNESTEN, Rabat, Morocco
} \\
{\small $^{2}$ High Energy Physics and Astrophysics Laboratory, FSSM, Cadi
Ayyad University, Marrakesh, Morocco } \\
{\small $^{3}$ Universit\'{e} Ibn Tofail, Facult\'{e} des Sciences, D\'{e}%
partement de Physique, LHESIR, K\'{e}nitra, Morocco.} }
\date{August 13, 2012}
\maketitle

\begin{abstract}
Considering the cosmological constant $\Lambda$ as a thermodynamic pressure
and its conjugate quantity as a thermodynamic volume as proposed in \cite{KM}%
, we discuss the critical behavior  of charged AdS black hole in
arbitrary dimensions $d$. In particular, we present a comparative
study  in terms of the spacetime dimension $d$ and the
displacement of critical points controlling the transition between
the small and the large black holes. Such behaviors vary nicely in
terms of $d$. Among our result in this context consists in showing
that the equation of state for a charged RN-AdS black hole
predicts an universal number given by $\frac{2d-5}{4d-8}$. The
three dimensional solution is also discussed.
\\
\\
\\
\\
\\
Few weeks after the submission of this work to Chinese Physics
Letters (CPL) on August 15th, we learned that authors of [arxiv:
1208.6251] found similar results to ours, more notably the
derivation of the same universal number given by
$\frac{2d-5}{4d-8}$.
\end{abstract}

\newpage The research in theories of black holes has scored remarkable
progress and has stimulated several interests. The spot is essentially
projected on the strength of such black holes in establishing the very deep
and fundamental relationship between the (quantum) gravity and the
thermodynamics more notably in anti-De Sitter Space. The principal key of
this relationship comes from the thermodynamic behavior of black holes,
where it appears that some laws of black holes are, in fact, simply the
ordinary laws of thermodynamics applied to a system containing a black hole.
We refer for instance to \cite{30}.

Recently, several efforts have been devoted to study phase transition and
critical phenomena for various AdS black hole backgrounds including
Reissner-Nordstrom AdS (RN-AdS) black solutions \cite{30,4,5,50,6,7,8}. In
particular, the $P=P(V,T)$ equation of state for a rotating black hole has
been dealt with and it has been found that this analysis share similar
feature as the Van der Waals $P-V$ diagram \cite{8,KM}.

More recently, these efforts have been pushed further by studying the P-V
criticality of RN-AdS black holes with spherical configurations. It has been
discussed the behavior of the Gibbs free energy in the fixed charge
ensemble. The authors of \cite{KM} have also reported on the phase
transition in the (P,T)-plane. In particular, it has been given a nice
interplay between the behavior of the RN-AdS black hole system and the Van
der Waals fluid. More precisely, P-V criticality, Gibbs free energy, first
order phase transition and the behavior near the critical points are
identified with the liquid-gas system.

The aim of this work is to contribute to these topics by studying such
behaviors in arbitrary dimensions. Applying an analogous analysis to \cite%
{KM} for the four dimensional case, we reconsider the critical behavior of
charged RN-AdS black holes in arbitrary dimensions of the spacetime.
Identifying, the cosmological constant $\Lambda$ as a thermodynamic pressure
and its conjugate quantity as a thermodynamic volume, we discuss such
behaviors by giving a comparative study in terms of the dimension and the
displacement of critical points. These parameters control the transition
between the small and the large black holes. It has been shown also that
such behaviors vary nicely in terms of the dimension of the spacetime in
which the black hole lives. Our computations predict an universal number
given by $\frac{2d-5}{4d-8}$. Then, a particular emphasis put on three
dimensional case corresponding to BTZ black hole.

To start let us first consider the Einstein-Maxwell-anti-de Sitter action in
higher dimensions $d$. The latter reads as
\begin{equation}
\mathcal{I}=-\frac{1}{16\pi G}\int_{M}dx^{d}\sqrt{-g}\left[ R-F^{2}+2\Lambda %
\right]   \label{action}
\end{equation}%

where the field strength $F$ is a closed 2-form. Indeed, it can always be
locally written as $F=dA$, where $A$ is the potential 1-form. In $d$
dimensions, $\Lambda $, identified with $-\frac{(d-1)(d-2)}{2\ell ^{2}}$, is
the cosmological constant associated with the characteristic length scale $%
\ell $. Varying the above action with respect to the metric tensor leads to
the RN-AdS solution given by
\begin{equation}
ds^{2}=-V(r)dt^{2}+\frac{dr^{2}}{V(r)}+r^{2}\Omega _{d-2}^{2}  \label{metric}
\end{equation}%
where $d\Omega _{d-2}^{2}$ indicates the metric on the $(d-2)$-dimensional
sphere. In this expression, the function $V(r)$ takes the following general
form
\begin{equation}
V(r)=1-\frac{2M}{r^{(d-3)}}+\frac{Q^{2}}{r^{2(d-3)}}+\frac{r^{2}}{\ell ^{2}}.
\end{equation}%
It is recalled that the parameter $M$ indicates the ADM mass of the black
hole while the parameter $Q$ represents the total charge. Using the
thermodynamic technics, we calculate the $d$ dimensional black hole
temperature. It is given by
\begin{equation}
T=\frac{1}{\beta }=\frac{1}{4\pi r_{+}}\left( (d-3)+\frac{(d-1)}{\ell ^{2}}%
r_{+}^{2}-\frac{(d-3)Q^{2}}{r^{2(d-3)}}\right)   \label{temperature}
\end{equation}%
where $r_{+}$ is the position of the black hole event horizon determined by
solving the condition $V(r_{+})=0$. In $d$ dimension, the entropy, as
usually, is given by
\begin{equation}
S\sim \frac{A}{4},  \label{entropy}
\end{equation}%
where now $A=\frac{2\pi ^{\frac{d-1}{2}}r^{d-2}}{\Gamma \left( \frac{d-1}{2}%
\right) }$. The electrical potential $\Phi $ measured at infinity with
respect to horizon takes the following form
\begin{equation}
\Phi =\frac{1}{c}\frac{Q}{r_{+}^{d-3}}  \label{phi}
\end{equation}%
where $c$ is a constant depending on the dimension $d$. It is given by $c=%
\sqrt{\frac{2(d-3)}{d-2}}$.

Using a similar analysis given in \cite{KM}, we can get the equation of
state for a charged AdS black hole $P=P(V,T)$ in arbitrary dimension $d$.
For a fixed charge value, the computation leads to
\begin{equation}  \label{state}
P=\frac{(d-2) T}{4 r_+}-\frac{(d-3) (d-2)}{16 \pi r_+^2}+\frac{(d-3) (d-2)
Q^2 r_+^{4-2 d}}{16 \pi }
\end{equation}
where $T$ is the temperature of the black hole. In this expression, the
event horizon radius $r_+$ takes the following general form
\begin{equation}  \label{hor}
r_+=\left(\frac{ \Gamma \left(\frac{d+1}{2}\right)V}{\pi ^{\frac{d-1}{2}}}%
\right)^{\frac{1}{d-1}}
\end{equation}
It is clear that the thermodynamic volume $V$ is related to the event
horizon radius $r_+$. It is recalled that $\Gamma$ is the gamma function.
For $d=4$, we recover the value given in \cite{KM}.

Mimicking the four dimensional analysis given in \cite{KM}, the physical
pressure and temperature become respectively as
\begin{equation}
Press=\frac{\hbar c}{l_P}P, \;\;\; Temp=\frac{\hbar c}{k} T
\end{equation}
where the Planck length is given by $l_P^2=\frac{\hbar G_d}{c^3}$. The
general expression, we are looking for, can be obtained by multiplying (\ref%
{state}) by $\frac{\hbar c}{l_P}$. Indeed, straightforward calculations lead
to the following form
\begin{equation}
Press=\frac{\hbar c}{l_P} \frac{(d-2) T}{4 r_+}+ \ldots =m\frac{k}{2l_P^2
r_+ }+\ldots
\end{equation}
A close inspection around the Van der Waals equation given by $\left(P+\frac{%
a}{v^2}\right)(v-b)=k T$ reveals that one can identify the specific volume $v
$ with the following expression
\begin{equation}
v=\frac{4 l _p^{2-d}}{(d-2)} r_+.
\end{equation}
In $d$ dimensions, the equation of state (\ref{state}) reads
\begin{equation}
P=\frac{T}{\mathit{v}}-\frac{d-3}{\pi (d-2) \mathit{v}^2}+\frac{ (d-3)
(d-2)^{(5-2d)} Q^2 }{16^{(3-d)}\pi \mathit{v}^{2(d-2)}}
\end{equation}
Numerical computations drive us to the corresponding "P-V" diagram, plotted
in figure1
\begin{center}
\begin{figure}[tbp]
\begin{tabbing}
\hspace{8cm}\=\kill
\includegraphics[scale=0.8]{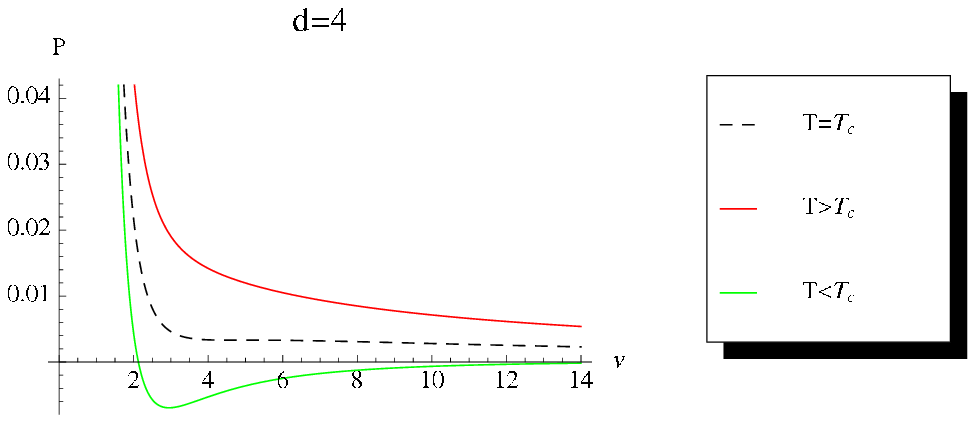}
\>\includegraphics[scale=0.8]{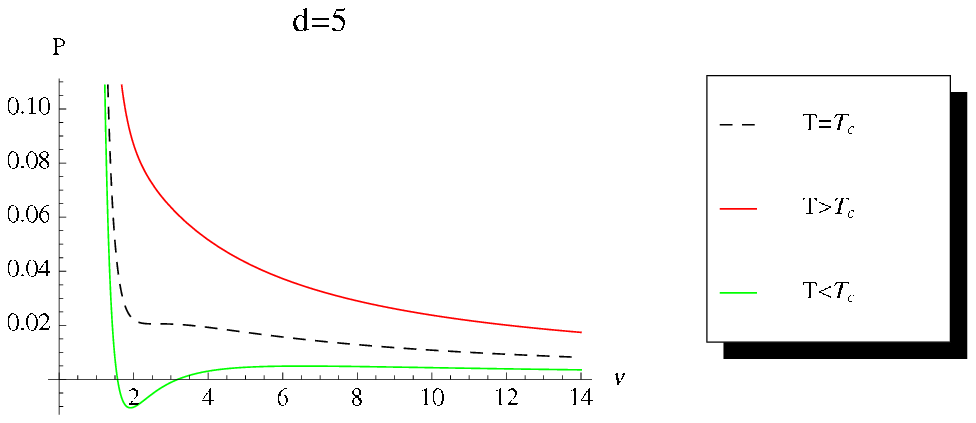} \\
\includegraphics[scale=0.8]{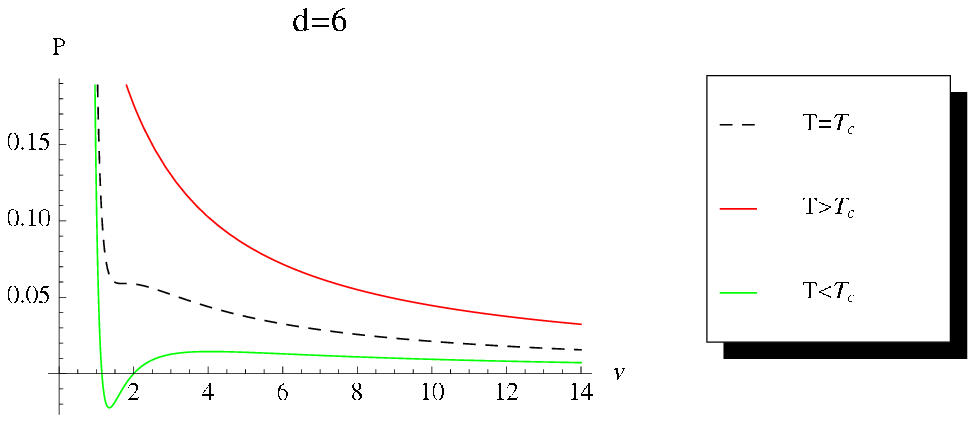}  \> \includegraphics[scale=0.8]{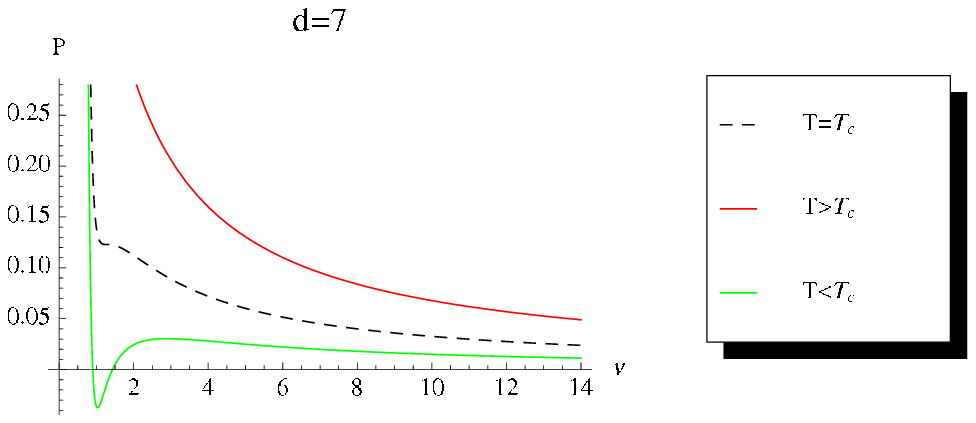}\\
\includegraphics[scale=0.8]{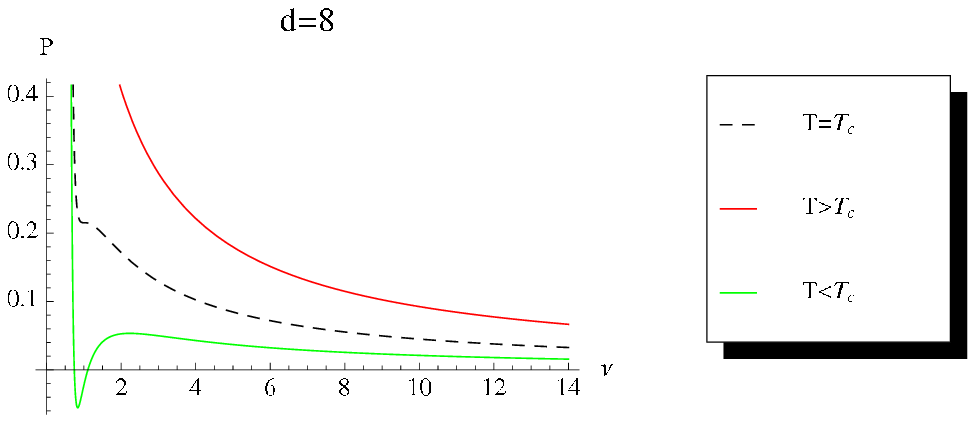}  \>\includegraphics[scale=0.8]{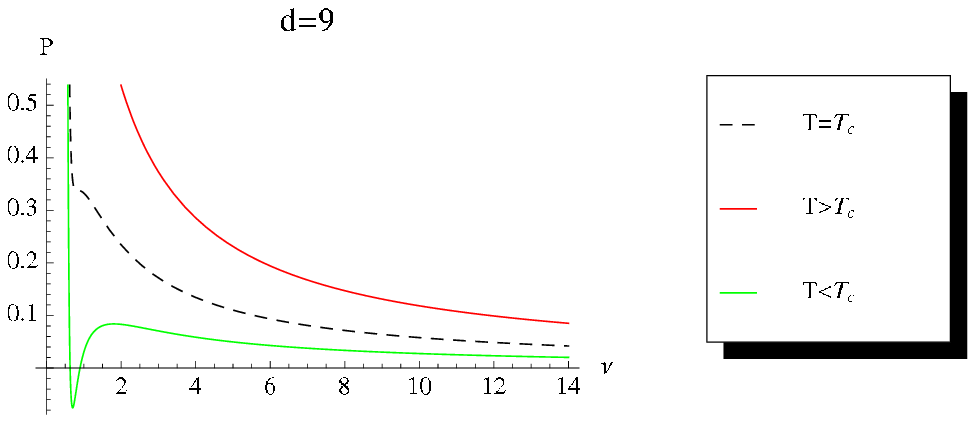} \\
\includegraphics[scale=0.8]{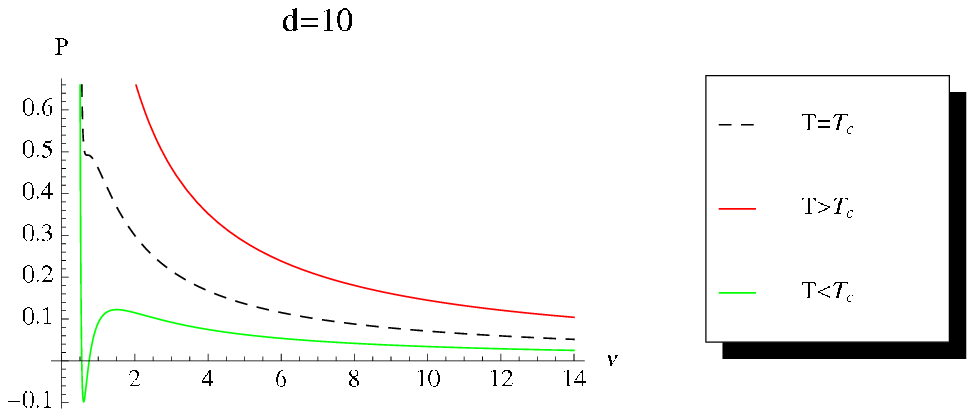}\>\includegraphics[scale=0.8]{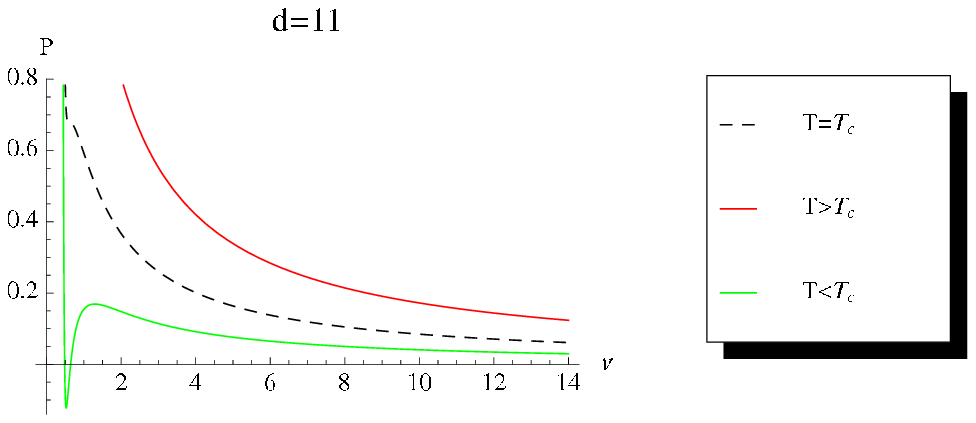}
\end{tabbing}
\vspace*{-.2cm} \caption{The $P-V$ diagram of charged AdS black
holes in arbitrary dimensions, where $T_c$ is the critical
temperature and the charge is equal to $1$ } \label{fig1:fig2}
\end{figure}
\end{center}
\newpage It follows that for $Q \neq 0$ and for $T< T_c$, the behavior looks
like an extended Van der Waals gas and the corresponding system involves
inflection points. As in four dimensions, the critical points are solution
of the following conditions
\begin{equation}
\frac{\partial P}{\partial v}=0,\;\;\;\; \frac{\partial^2 P}{\partial v^2}=0.
\end{equation}
At the vicinity of such critical values, we have calculated the following
thermodynamic quantities

\begin{center}
\begin{tabular}{|l|c|c|c|c|r|r|}
\hline
\textbf{\textit{$d$ }} & $T_c$ & $v_c$ & $P_c$ & $cst$   \\ \hline\hline
\textbf{\textit{4}} & $\frac{1}{3 \sqrt{6} \pi Q}$ & $2 \sqrt{6} Q$ & $\frac{%
1}{96 \pi Q^2}$ & $\frac{3 }{8} $   \\ \hline
\textbf{\textit{5}} & $\frac{4}{5 \sqrt[4]{15} \pi \sqrt{Q}}$ & $\frac{4
\sqrt[4]{5} \sqrt{Q}}{3^{3/4}}$ & $\frac{1}{4 \sqrt{15} \pi Q}$ & $\frac{5 }{%
12}$   \\ \hline
\textbf{\textit{6}} & $\frac{9}{7 \sqrt[3]{2} \sqrt[6]{7} \pi \sqrt[3]{Q}}$
& $\sqrt[3]{2} \sqrt[6]{7} \sqrt[3]{Q}$ & $\frac{9}{16 2^{2/3} \sqrt[3]{7}
\pi Q^{2/3}}$ & $\frac{7 }{16}$   \\ \hline
\textbf{\textit{7}} & $\frac{16}{9 \sqrt[4]{3} \sqrt[8]{5} \pi \sqrt[4]{Q}} $
& $\frac{4 \sqrt[4]{3} \sqrt[4]{Q}}{5^{7/8}}$ & $\frac{1}{\sqrt{3} \sqrt[4]{5%
} \pi \sqrt{Q}}$ & $\frac{9 }{20} $   \\ \hline
\textbf{\textit{8}} & $\frac{25}{11 \sqrt[10]{66} \pi \sqrt[5]{Q}}$ & $\frac{%
2 \sqrt[10]{22} \sqrt[5]{Q}}{3^{9/10}}$ & $\frac{25}{16 \sqrt[5]{66} \pi
Q^{2/5}}$ & $\frac{11 }{24}$  \\ \hline
\textbf{\textit{9}} & $\frac{36}{13 \sqrt[12]{91} \pi \sqrt[6]{Q}}$ & $\frac{%
4 \sqrt[12]{13} \sqrt[6]{Q}}{7^{11/12}}$ & $\frac{9}{4 \sqrt[6]{91} \pi \sqrt%
[3]{Q}}$ & $\frac{13 }{28}$   \\ \hline
\textbf{\textit{10}} & $\frac{49}{15 2^{3/14} \sqrt[14]{15} \pi \sqrt[7]{Q}}$
& $\frac{\sqrt[14]{15} \sqrt[7]{Q}}{2^{11/14}} $ & $\frac{49}{16 2^{3/7}
\sqrt[7]{15} \pi Q^{2/7}} $ & $\frac{15 }{32}$   \\ \hline
\textbf{\textit{11}} & $\frac{64}{17 \sqrt[8]{3} \sqrt[16]{17} \pi \sqrt[8]{Q%
}}$ & $\frac{4 \sqrt[16]{17} \sqrt[8]{Q}}{3 3^{7/8}}$ & $\frac{4}{\sqrt[4]{3}
\sqrt[8]{17} \pi \sqrt[4]{Q}} $ & $\frac{17 }{36} $  \\ \hline
\end{tabular}
\label{critic}
\end{center}

It is observed that the quantity $\frac{P_c v_c}{T_c}$ varies when the
dimension of the black hole is changed. Here, we are very interested in how
this change looks like. It is recalled that  for $d=4$, this quantity is
fixed to $\frac{3}{8}$ being exactly the same for the Van der Waals fluid
and it is considered as an universal number predicted for any charged AdS
black hole. We will see that such a number behaves nicely in terms of $d$.
To see that, we first vary the dimension $d$, then we plot the corresponding
result in figure 2.

\begin{center}
\begin{figure}[!h]
\hspace{3 cm} {\includegraphics[scale=.9]{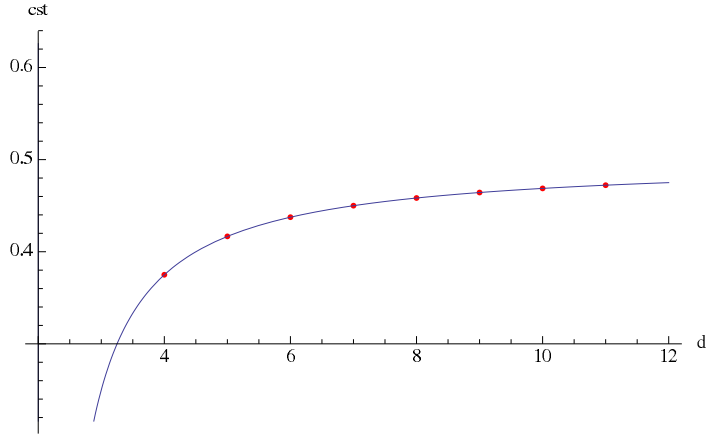}}
\vspace*{-.2cm} \caption{The behavior of the universal constant
$cst$ in terms of $d$. } \label{fig1:fig2}
\end{figure}
\end{center}

\newpage
 Fitting our result, the behavior of the extended universal quantity
takes a very nice form. Indeed, it is given by
\begin{equation}
\frac{P_c v_c}{T_c}=\frac{2d-5}{4d-8}
\end{equation}
This is a nice general expression since it gives, as a particular case for $%
d=4$, the value $\frac{3}{8}$ obtained in \cite{KM}. This shows that the
critical points controlling the transition between the small and the large
black hole vary in terms of the dimension of the spacetime. This numerical
calculation reveals that the localization of the critical points depends on
the dimension $d$. Augmenting the dimension o
f the spacetime such points get
dapper. Moreover, it has been observed that small black hole phase becomes
relevant for higher dimensional models.

We have to underline that it is remarkably seen that the extended universal
constant number $\frac{2d-5}{4d-8}$ share a striking resemblance with the
ideal gas constant $R$.

It is worth noting that the critical points obey a nice polynomial shape.
The obtained result is presented in figure 3

\begin{center}
\begin{figure}[!h]
\hspace{3 cm} {\includegraphics[scale=.9]{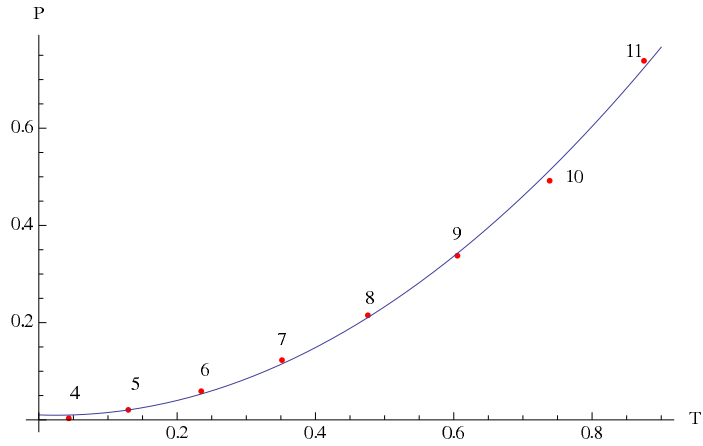}}
\vspace*{-.2cm} \caption{Critical point positions in function of
dimensions $d$ } \label{fig1:fig2}
\end{figure}
\end{center}

Fitting this numerical calculation, the behavior of such points is
controlled by the following polynomial function
\begin{equation}
P=\alpha+\beta T+\gamma T^2
\end{equation}
For a particular choice in the parameter space where the charge $Q$ is fixed
to 1, the above coefficients are given by
\begin{eqnarray}
\alpha=0.0102026, \qquad \beta=0.0492466\qquad \gamma=0.988634.
\end{eqnarray}
It will be interesting to understand the polynomial behavior of the critical
points in terms of the dimension.

In the end of this work, we would like to make comment on a particular case
corresponding to $d=3$. In this case, the equation (\ref{state}) reduces to
\begin{equation}
Pv=T
\end{equation}
describing an ideal gas equation. Treating the dimension as a size
parameter, it seems that the Van der Walls analysis can be converted to an
ideal gas model. On the other hand, it is recalled that in  $d=3$ lives a
BTZ black hole solution\cite{09}. Following \cite{9,90}, the BTZ temperature
is given by
\begin{equation}
T=\frac{1}{4\pi}\left(\frac{2r_+}{\ell^2}-\frac{\pi Q^2}{r_+}\right).
\end{equation}
The corresponding state equation can be written as
\begin{equation}
P=\frac{Q^2}{8 \pi r_+^2}+\frac{T}{4 r_+}.
\end{equation}
This looks like (\ref{state}) without critical points. It follows that BTZ
black hole satisfies an ideal gas behavior. This is can be obtained by
turning off the coupling Maxwell gauge fields.

Note by the way, that the state equation corresponding to a $d$ dimensional
BTZ black hole reads as
\begin{equation}
P=\frac{(d-2) T}{4 r_+}+\frac{2^{\frac{d-9}{2}} (d-2) Q^{d-1} r_+^{1-d}}{\pi
}.
\end{equation}

Finally, let us mention that the above behavior present only $d>2$. For $d=2$%
, the equation of state becomes $P=0$. It should be interesting to
reconsider the discussion of such a case in connection with $2d$ black holes
and Liouville theory. This will be addressed in future works.

\end{document}